\documentclass[preprint,showpacs,preprintnumbers,amssymb]{revtex4}

\begin{document}
\title{Re-analysis of the $D^* D\pi$ coupling in the light-cone QCD 
sum rule}
\author{Hungchong Kim}
\email{hung@phya.yonsei.ac.kr}
\affiliation{%
Institute of Physics and Applied Physics, Yonsei University,
Seoul 120-749, Korea
}
\begin{abstract}
The recent measurement from the CLEO experiment presents  
the $DD^*\pi$ coupling, $17.9\pm 0.3 \pm 1.9$. This value is
much larger than any of QCD sum rule predictions available
in literature.  We report that, with a relevant treatment of
the continuum subtraction as well as with the asymptotic form
of the twist-2 pion wave function, the light-cone QCD sum rule
can provide the coupling comparable to the experimental value.
The stability of the resulting sum rule becomes much better
with these corrections.
\end{abstract}

\maketitle

\section{Introduction}
\label{sec:intro}

Recently, the CLEO collaboration~\cite{Anastassov:2001cw} presents 
a new measurement of the
$D^*D\pi$ coupling, $g_{D^*D\pi} =
17.9\pm 0.3 \pm 1.9$. Though this value is somewhat consistent with
quark model predictions~\cite{Becirevic:1999fr} and
lattice calculations~\cite{Abada:2002xe},
it is much larger than any of QCD sum rule predictions, which gives
some skepticism in accepting the various QCD sum rule results.
In particular,
the conventional QCD sum rules relying on the short-distance
expansions~\cite{Colangelo:1994es,Colangelo:1995ph,DiBartolomeo:1995ir,
Kim:2001es} consistently restrict the coupling $g_{D^*D\pi} \le 10$.
In these sum rules, QCD inputs are well-constrained by the low-energy
theorem.  Thus, the current disagreement may due to the
fact that various pion matrix elements appearing in the
conventional sum rules may not converge fast enough in the short-distance
expansions. One may need to do a partial summation to all order of those
matrix elements.

The other method is to construct QCD sum rules from three-point
function~\cite{Navarra:2000ji}, which leads to the coupling around 6,
again much smaller than the experimental value. 
Its recent extension~\cite{Navarra:2001ju} using exponential type 
parametrization for the form factor
seems to give a much larger value around 15.  However, it should be remembered
that the operator product expansion (OPE) of a three-point function leads to an
unphysical behavior at high momentum transfers~\cite{Braun:1997kw}, which
in fact was one of the motivation for constructing light-cone QCD sum 
rules (LCQSR) relying on the expansions along the light-cone.
Also, LCQSR may improve the conventional QCD sum rules by
having the pion wave functions that encode the partial summation 
to all order of the pion matrix elements.
However, LCQSR~\cite{Belyaev:1995zk,Colangelo:1998rp} yields 
the $D^*D\pi$ coupling only about 12, still smaller than
the experimental value. Thus, as far as the $D^*D\pi$ coupling is
concerned, all the sum rules do not 
provide the experimental value.  Furthermore as 
quark model and lattice calculations~\cite{Becirevic:1999fr, Abada:2002xe}
agree with the experiment, it is important to re-investigate
the existing sum rules and look for a way to reach an
agreement with them. 

One crucial input in the prediction of LCQSR is the twist-2 pion wave 
function at the middle point $\varphi_\pi(0.5)$.
Refs.\cite{Belyaev:1995zk,Colangelo:1998rp} in their
sum rules use
$\varphi_\pi(0.5)\sim 1.2$ obtained by comparing 
two different Lorentz structures of the light-cone sum rule for
the pion-nucleon coupling~\cite{Braun:1988qv}.
However, as the pion-nucleon sum rule
has a strong dependence on the Lorentz structure~\cite{Kim:1998ir}
considered, $\varphi_\pi(0.5)\sim 1.2$ is questionable.
Thus, one may attribute the current
discrepancy in the $D^*D\pi$ coupling to the uncertainty of $\varphi_\pi(0.5)$.
Indeed, there are some suggestions that the twist-2 pion wave function should
be of the asymptotic form whose value at the middle point is 
$\varphi_\pi(0.5)\sim 1.5$~\cite{Zhu:1998am,Belyaev:1997iu}. 
As we will discuss below, this new input only
increases the $D^*D\pi$ coupling by 9 \%, still not enough to
reproduce the experimental value.
Certainly, the existing sum rules need further improvements.

In this letter, we suggest one possibility to improve
the existing LCQSR. Specifically, we show that the continuum subtraction in
the existing LCQSR is mathematically ill-defined.
A similar suggestion was reported in 
Refs.~\cite{Kim:2000ku,Kim:2000rg,Kim:2000ur}.
In particular, we demonstrate that the OPE of the form
$\int^1_0 du/[m_c^2 -up_1^2 -(1-u)p^2_2]$ should not entail the continuum
subtraction if one strictly follows the QCD duality assumption 
in constructing the continuum contributions.
This correction combined with the asymptotic twist-2 pion wave function leads 
to the $D^*D\pi$ coupling
comparable to its experimental value.  The resulting sum rule is  
stable against the variation of the Borel mass.

\section{The continuum construction}
\label{sec:continuum}

In this section, we demonstrate a possible modification
in obtaining the continuum
subtraction in the existing LCQSR~\cite{Belyaev:1995zk,Colangelo:1998rp}.
The leading OPE in the existing LCQSR 
takes the form
\begin{eqnarray}
\int^1_0  {du \over m_c^2 -up_1^2-(1-u)p_2^2} 
\label{ope}
\end{eqnarray}
where 
$m_c$ the charm quark mass and $p_1,p_2$ are the two
momenta associated with the coupling vertex.
The leading OPE in Refs.~\cite{Belyaev:1995zk,Colangelo:1998rp} also
contains the pion wave function in the numerator of the integral but here we
suppress that for a mathematical simplicity.   
The main issue regarding the continuum subtraction 
is not affected by this simplicity.

It was claimed in Ref.\cite{Belyaev:1995zk,Colangelo:1998rp} 
that, after the continuum subtraction and the double Borel transformation,
the OPE, Eq.(\ref{ope}), appears in the final sum rule as the
factor
\begin{eqnarray}
M^2 (e^{-m_c^2/M^2} - e^{-S_0/M^2})
\label{simple}
\end{eqnarray}
where $M^2$ is the reduced mass of the two Borel
masses associated with the double Borel transformations.
We will show that the continuum subtraction factor,
$M^2e^{-S_0/M^2}$, {\it comes from a mathematically spurious 
term and it should not be a part of
the final sum rule}.

To determine the continuum subtraction for the given OPE, one needs to
determine the spectral density $\rho(s_1,s_2)$ firstly from the double 
dispersion relation,
\begin{eqnarray}
\int^1_0 {du \over m_c^2 -up_1^2-(1-u)p_2^2}=
\int^\infty_0 ds_1 \int^\infty_0 ds_2 {\rho(s_1,s_2) \over 
(s_1-p_1^2)(s_2-p_2^2)}\ .
\label{double}
\end{eqnarray}
Note that the lower boundaries of the integrals in the right-hand side 
should be chosen
such a way that the entire region of $\rho(s_1,s_2) \ne 0$ is
included.
According to the prescription given in Ref.\cite{Belyaev:1995zk}, 
the spectral density is given by
\begin{eqnarray}
\rho(s_1,s_2)= \delta(s_1-s_2) \theta(s_1-m_c^2).
\label{spec}
\end{eqnarray}
An alternative way to obtain this spectral density is to take the
imaginary part of Eq.(\ref{ope}) with respect to the
two momenta, $p_1^2$ and $p_2^2$.

A common practice in LCQSR is to (1) put this spectral density back 
to the double dispersion integral,
(2) restrict the integral below the continuum threshold $S_0$ (according
to QCD duality), $\int^\infty_0 ds_1 \int^\infty_0 ds_2\rightarrow 
\int^{S_0}_0 ds_1 \int^{S_0}_0 ds_2$,
and (3) take the double Borel transformations.  This procedure 
as advocated by Belyaev {\it et.al.}~\cite{Belyaev:1995zk} 
precisely yields the simple prescription given in Eq.~(\ref{simple}).
We want to point out that the second
step is dangerous as it yields a spurious contribution.
To show this, let us see how 
the spectral density Eq.(\ref{spec}) reproduces the OPE Eq.(\ref{ope})
within the double dispersion
relation Eq.(\ref{double}). 
We put the spectral density Eq.(\ref{spec}) in the 
double dispersion relation and  obtain
\begin{eqnarray}
\int^\infty_0 ds_1 \int^\infty_0 ds_2 {\rho(s_1,s_2) \over 
(s_1-p_1^2)(s_2-p_2^2)} = \int^1_0 du \int^\infty_0 ds_1 
{\theta(s_1-m_c^2)\over (s_1 -up_1^2 -(1-u)p_2^2)^2}\ .
\label{dis}
\end{eqnarray}
Note, we have used the Feynman parameterization to get
an integral over the Feynman parameter $u$.
We then perform the integration by part to separate into the
two terms 
\begin{eqnarray}
\int^1_0 du \int^\infty_0 ds_1 {\delta(s_1-m_c^2)\over s_1 -up_1^2 -(1-u)p_2^2}
-\int^1_0 du 
{\theta(s_1-m_c^2)\over s_1 -up_1^2 -(1-u)p_2^2}\Bigg |^{s_1=\infty}_{s_1=0}\ .
\label{dis2}
\end{eqnarray}
One can see that the first term precisely yields the anticipated OPE while
the second term, though vanishes under the present boundaries,  
is an unnecessary spurious term.  
When QCD duality is applied ({\it i.e.} restricting the integral
below the continuum threshold $S_0$),
it is mathematically sensible to apply 
to {\it the first term only}. Any contribution from the second term
is ill-defined as the second term is mathematically 
spurious.  
However one can show that the continuum subtraction factor
$M^2e^{-S_0/M^2}$ comes from {\it the spurious second term}.
When the integral interval in Eq.~(\ref{dis}) is switched to $0\sim S_0$,
the second term becomes
$\int^1_0 du /[S_0 -up_1^2 -(1-u)p_2^2]$, which no longer vanishes. Then   
under the double Borel transformations  the second term
precisely yields the continuum subtraction
factor $M^2e^{-S_0/M^2}$.  Therefore,  
the continuum subtraction factor $M^2e^{-S_0/M^2}$ should be dropped in
the final expression of the sum rule 
as it is from the mathematically spurious term.

Once the factor $M^2e^{-S_0/M^2}$ is dropped,
then the resulting sum rule does not depend on the continuum threshold,
indicating that there is no continuum contribution.
Intuitively, absence of the continuum contribution
may seem strange because the current in the
correlation function can couple to higher resonances as well
as the lowest resonance. But one can show that $\alpha_s$ corrections
to the perturbative part can give a continuum contribution 
within our prescription 
as the corrections are logarithmic functions of $s_1$ and 
$s_2$~\cite{Khodjamirian:1999hb}. 
Therefore one can pick up small continuum contribution 
[at an order $\cal{O}$($\alpha_s$)] and, in the
present calculation without $\alpha_s$ correction, having no 
continuum contribution is not against the intuitive picture.

\section{Re-calculation of the $D^*D\pi$ coupling}
\label{sec:analysis}
\begin{figure}
\caption{$f_D f_{D^*} g_{D^* D\pi}$ versus the Borel mass in
the light-cone QCD sum rule. The solid line is from
the sum rule of Ref.~\cite{Belyaev:1995zk}.  The dash curve is
obtained when $\varphi_\pi(0.5)=1.5$ is used. The dot-dashed curve is
obtained with the continuum correction.
}
\label{fig1}

\setlength{\textwidth}{6.1in}   
\setlength{\textheight}{9.in}  
\centerline{%
\vbox to 2.4in{\vss
   \hbox to 3.3in{\includegraphics{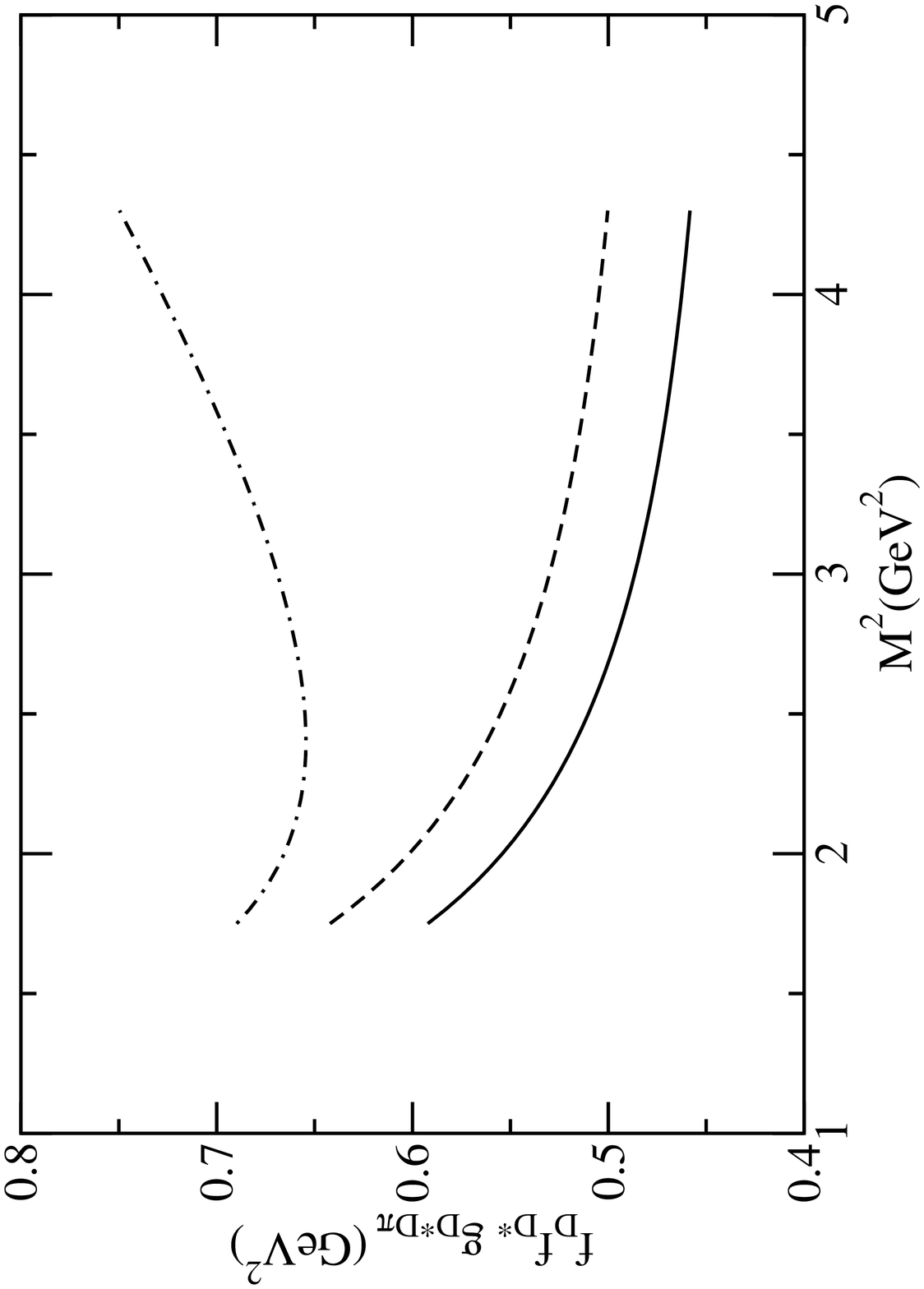}\hss}}
}

\vspace{50pt}
\end{figure}

Having suggested a modification in the
previous LCQSR, we now re-analyse the sum rule
for the $D^*D\pi$ coupling.  We take the sum rule formula for the
$D^*D\pi$ coupling from Ref.\cite{Belyaev:1995zk} and obtain
the solid curve in Fig.~\ref{fig1}.   
This is the same curve that was presented 
in Ref.\cite{Belyaev:1995zk}.  Based on this curve, 
Ref.\cite{Belyaev:1995zk} obtained $f_D f_{D^*} g_{D^*D\pi}=0.51$ GeV$^2$,
which yields $g_{D^*D\pi}=12.5$ if we use $f_D=170$ MeV and $f_{D^*}=240$ 
MeV obtained from two-point vacuum sum rules.  
The dashed curve is obtained when we take the asymptotic twist-2 pion
wave function, $\varphi_\pi(0.5)=1.5$, while keeping all other
parameters fixed as in Ref.\cite{Belyaev:1995zk}.  
This correction increases the coupling
only by 9 \%. Thus, with this correction only, LCQSR cannot still 
reproduce the experimental value of the coupling, 17.9. 

However the correction coming from the continuum factor is substantial.
The dot-dashed curve is obtained when the continuum subtraction factor
$M^2 e^{-S_0/M^2}$
is taken out in the sum rule according to our suggestion in
Sec.\ref{sec:continuum}.  One clearly sees that this curve can provide
a much larger coupling.  Furthermore, the curve has a minimum around which 
the variation with respect to the Borel mass is minimized. 
This indicates that, around the minimum,  the sum rule is well 
saturated by the OPE terms included in the calculation. 
For a qualitative estimate, 
instead of doing a detail analysis, we take the minimum of the curve to
calculate the coupling. The minimum gives $f_D f_{D^*} 
g_{D^*D\pi}=0.66$ GeV$^2$,
which in turn yields the coupling $g_{D^*D\pi}=16.2$. 
This value is only 10 \% smaller
than the experimental one.  
The other source of error is the uncertainty in the decay constants
$f_D$ and $f_{D^*}$.  Indeed, from Ref.\cite{Belyaev:1995zk}, 
the decay constants are known to be $f_D = 170\pm 10$ MeV
and $f_{D^*} = 240 \pm 20$ MeV.  If we take these uncertainties into
account we obtain the range 
\begin{eqnarray}
14.1 \le g_{D^* D\pi} \le 18.75
\end{eqnarray} 
which certainly overlaps with the experimental value.

Of course, there are other possibilities
to improve the existing LCQSR. For example,
the current LCQSR is up to twist-4.  There might be some
contributions from higher twists. Also the twist-4 element
denoted by $\delta^2$ is currently known to be $0.2$ GeV$^2$
based on QCD sum rule calculation of Novikov {\it et. al.}~\cite{Novikov:jt}
This value however crucially depends on the factorization assumption
for the four-quark condensate $\langle ({\bar q} q)^2 \rangle$. 
Based on the rho meson sum rule~\cite{Leupold:2001hj}, the four-quark
condensate can be much larger, which may shift $\delta^2$ to
a larger value.
Nevertheless, what we want to emphasize 
is that all these improvements are expected to be less than 10 \% because,
otherwise, the twist expansion can not be valid.
To be comparable with the CLEO experiment, one needs a correction of order
40 \% in the existing LCQSR calculation. Therefore, 
the continuum correction that we are addressing in this work
provides the most important modification to the existing
LCQSR and it helps to reach an agreement with the CLEO experiment.
More interesting is that, with this improvement, 
LCQSR provides the coupling comparable to the quark 
model~\cite{Becirevic:1999fr} and 
lattice calculations~\cite{Abada:2002xe}.

\acknowledgments 
The work was supported by Korean Research Foundation Grant
(KRF-2002-015-CP0074).

\eject

\end{document}